\documentclass[11pt]{article}
\usepackage[utf8]{inputenc}
\usepackage{setspace}
\usepackage{palatino}
\usepackage{graphicx}
\usepackage{float}
\usepackage{titling} 
\usepackage{multirow}
\usepackage{lscape}
\usepackage{amsmath}
\usepackage{amssymb}
\usepackage{subcaption}
\usepackage[a4paper, total={6in, 9.5in}]{geometry}
\fontfamily{ppl}\selectfont 

\usepackage[american]{babel}

\usepackage{setspace}

\usepackage[natbibapa]{apacite} 
\usepackage{lipsum}

\usepackage{todonotes}
\usepackage{times}
\usepackage{soul}
\usepackage{url}
\usepackage[hidelinks]{hyperref}
\usepackage{cleveref}
\usepackage[utf8]{inputenc}
\usepackage{caption}
\usepackage{graphicx}
\usepackage{amsmath}
\usepackage{amsthm}
\usepackage{booktabs}
\usepackage{algorithm}
\usepackage{algorithmic}
\usepackage{comment}
\usepackage[switch]{lineno}
\usepackage{indentfirst}

\author{Haochen Li\\ { King's College London}\\
    {\texttt{ haochen\_li@kcl.ac.uk}} 
\and Maria Polukarov\\ { King's College London}\\
    {\texttt{maria.polukarov@kcl.ac.uk}}
\and Carmine Ventre\\ { King's College London}\\
    {\texttt{carmine.ventre@kcl.ac.uk}}
    }

\title{Detecting Financial Market Manipulation with Statistical Physics Tools}

\begin{document}
\begin{titlepage}

    \setcounter{footnote}{1} 
    \maketitle

    \begin{abstract}

        We take inspiration from statistical physics to develop a novel conceptual framework for the analysis of financial markets. We model the order book dynamics as a motion of particles and define the \emph{momentum} measure of the system as a way to summarise and assess the state of the market. Our approach proves useful in capturing salient financial market phenomena: in particular, it helps detect the market manipulation activities called spoofing and layering. We apply our method to identify pathological order book behaviours during the flash crash of the LUNA cryptocurrency, uncovering widespread instances of spoofing and layering in the market. Furthermore, we establish that our technique outperforms the conventional Z-score-based anomaly detection method in identifying market manipulations across both LUNA and Bitcoin cryptocurrency markets.

    \end{abstract}


\end{titlepage}

\newpage
\section{Introduction}
%

In the last decade, we have seen a large increase in investments into decentralised digital markets, characterised by attractively low transaction fees and the prospect of high profits. 
However, due to their relatively unregulated nature and the large proportion of private retail investors and non-institutional traders, these markets are highly susceptible to 
market manipulation.  
It is therefore of critical importance to have tools to effectively detect such activities. 
To this end, in this paper we define and evaluate a statistical physics approach for analysing the 
order dynamics of financial markets.  

\smallskip \noindent {\bf Contribution.} A market is typically organised via a central record, called the \emph{limit order book} (LOB), of all the orders submitted by the traders. To capture the full LOB market data (a.k.a., Level 3 data), 
we model the orders as physical particles, with order submissions as particles incoming the system, order cancellations as particles exiting the system, and each order transaction as a particle annihilation. We can then calculate the change of \emph{momentum} in the system to estimate the impact of each action within. In essence, our method 
uses the microscopic dynamics of the order book  activities to build a systemic assessment of the market behaviour and visually highlight the events happening therein. 
To the best of our knowledge, there is no other analytical tool that makes use of the tremendous amount of data in this domain, to effectively illustrate the financial market phenomena. To demonstrate the power of our approach, we show it can successfully 
detect 
spoofing and layering, two of the most common market manipulation techniques. 

Specifically, we evaluate our framework in the context of the May 2022 flash crash of the LUNA cryptocurrency. Looking at the LOB data corresponding to the LUNA/USD pair, we can easily see how the \emph{limit} and \emph{market} orders were pushing the market in opposite directions. Indeed, limit orders are less aggressive and set a \emph{limit} price to transact, whereas market orders buy or sell at the current \emph{market} price and are then much more aggressive. During the flash crash of LUNA, the market orders endeavour to elevate the price without success, whereas limit orders consistently engage in selling activities. This observation indicates that the decline in LUNA's price is predominantly attributable to the prevalence of limit orders, which effectively counteract the strength of market orders. Furthermore, we find evidence that spoofing and layering were widely present in the market at the time. We further apply our method to analyse the Bitcoin trading market and ascertain its enhanced efficacy in identifying market manipulation endeavours compared to conventional Z-score-based approaches. 


\smallskip \noindent {\bf Related work.} Here we overview the literature 
on detecting market manipulations and 
the application of physical models to LOBs.

\cite{tuccella2021protecting} presented a Gated Recurrent Unit model 
to detect and identify spoofing events on several cryptocurrencies 
using (the slightly less rich) Level 2 data. \cite{leangarun2016stock} produced 1-min LOB snapshots from Level 2 data to train neural networks for detecting spoofing in NASDAQ stock markets. These methods use supervised learning and thus require labelled data. However, in electronic markets there are various patterns of market manipulations that depend on particular market conditions. 
As a result, it is not feasible to systematically and effectively label all the 
occurrences of market manipulation.  
In the context of unsupervised anomaly detection, the traditional method is to specify the time series as the base 
and decide whether a newly occurred observation abnormally deviates from the base behaviour according to \cite{gunnemann2014robust}. However, these attempts used time series data leading to the loss of temporal information that are key in spoofing detection.  
In contrast, our 
method 
uses all of the order records without losing any temporal information. 

\cite{cartea2020spoofing} investigated the imbalance between the bid and ask volumes whereas \cite{mendoncca2020detection} produced 1-min LOB snapshots from Level 2 order book data to detect spoofing in Brazilian stock markets. These papers, again, do not take full advantage of the Level 3 data, 
thus 
losing information. Moreover, neither of these works could 
visualise the LOB and the pernicious orders. 

\cite{lillo2003master} observed how  
the oscillations from the supply-and-demand equilibrium for various financial assets 
are controlled by the equivalent statistical law. They concluded that it was helpful to model human activity as random rather than rational under certain circumstances. 
%
This and the follow-up work \cite{farmer2005predictive,lillo2005key} 
proved that LOBs are effectively 
stochastic processes, hence there are certain statistical rules thereon. 
Thus, it is feasible to study the rules on the activities of these orders in the context of statistical physics.

\cite{yura:2014financial,yura:2015financial} treated the orders placed in foreign exchange markets as fluid particles with diameter equivalent to the minimum trading unit, and distinguished the order book to different layers according to a concept called interaction range defined by the activity of orders. Therefore, they modelled the bid-ask spread and the nearest price ticks with temporarily largest volume as a colloidal financial Brownian particle. 

\section{Model}
\label{Model}

In this section, we introduce the physical model to describe the limit order book and the order activity.

We denote the market highest buy (best bid) and lowest sell (best ask) quoted prices in the LOB with $b_M$ and $a_M$, respectively. The market midprice is defined as $z_M=\frac{b_M+a_M}{2}$. Every time a trade occurs, we also record the market transacted (or, match) price, $p$.

At time $t$, the market midprice is denoted by $z_M(t)$, and the number of limit orders at the quoted price $x$ is denoted by $N(x,t)$ which is always non-negative. For the time period $[t-\Delta t, t]$, the \emph{velocity} of the midprice is given by the change in the midprice per unit of time: $v_M(t)=\frac{z_M(t)-z_M(t-\Delta t)}{\Delta t}$.

\begin{figure}[!ht]
\begin{center}
\includegraphics[width=0.9\columnwidth]{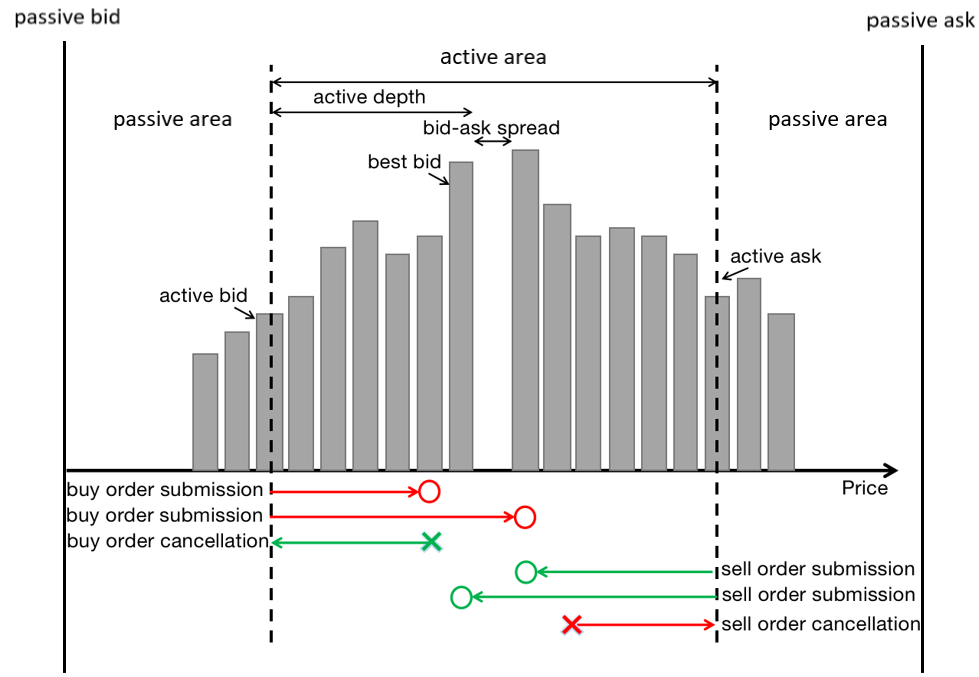}
\end{center}
\caption{Active and passive areas of the order book}
\label{fig:foo2}
\end{figure}

The order activities tend to occur more frequently around the market midprice. 
An \emph{active area} is a wider range at a specific depth in the order book, which  is centred on the bid-ask spread, see Figure~\ref{fig:foo2}. Intuitively, the active area follows the movement of the midprice: Inside this range the order submissions and cancellations are the most active, whilst outside this range the limit orders tend to keep static. The \emph{active bid (sell) price} is then defined as the best bid (ask) price minus (plus) the depth of the active area. 

\begin{figure}
\centerline{\includegraphics[height=2in]{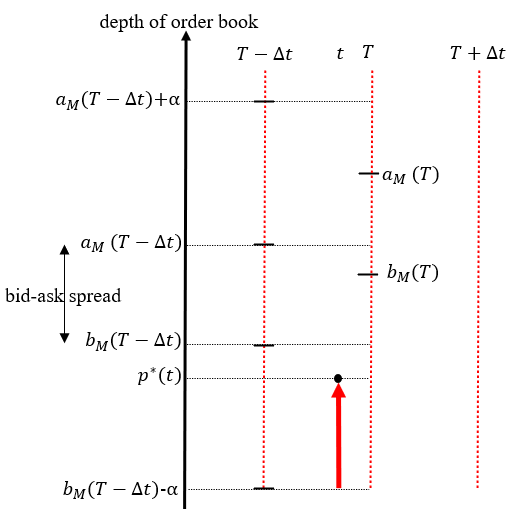}}
\caption{Motion of a limit buy order} \label{fig:foo3}
\end{figure}

We view the orders as physical particles with motion on a 1-dimensional axis. The order size corresponds to the particle mass, and the distance the order has moved corresponds to the distance the particle moves. 
We write the active area interval as $[b_M-\alpha, a_M+\alpha]$ where $\alpha$ is the active area's depth. 

Consider a  limit buy order of size $s$ submitted to a quoted price $p^*(t)$ at time $t$ during the sampling period $(T-\Delta t,T]$, $t \in (T-\Delta t,T]$. (Note that $p^*\le a_M$ as limit orders with a quoted price larger than the best ask price $a_M$ will be matched to execute at price $a_M$ so that $p^*=a_M$.) We then say that the order has moved from the depth of the lower bound of the active area $b_M-\alpha$ at time $T - \Delta t$ to its quoted price $p^*(t)$ at time $T$. As depicted in Figure~\ref{fig:foo3}, the limit buy order indicated by the red arrow has moved from the lower bound of the active area  located at $b_M(T-\Delta t)-\alpha$ (which will remain the same during the sampling period $(T-\Delta t,T]$) at time $T-\Delta t$ to location $p^*$ at time $t$ for a distance of $p^*(t)-(b_M(T-\Delta t)-\alpha)$. As a result, the velocity of this limit buy order is 
\begin{equation}
    v(t) = \frac{p^*(t)-(b_M(T-\Delta t)-\alpha)}{\Delta t}.
\label{eq:2}
\end{equation}
For limit buy orders with the quoted price higher than the market best ask, $a_M$, we have $p^*(t)=a_M(T-\Delta t)$. 

A buy order that is cancelled is treated as moving from its previous quoted price $p^*$ to the depth of the lower bound of the active area, $b_M-\alpha$, at time $t$, with the velocity of 
\begin{equation}
    v(t) = \frac{(b_M(T-\Delta t)-\alpha)-p^*(t)}{\Delta t}.
\label{eq:3}
\end{equation}
Similarly, the velocity of a limit sell order is given by 
\begin{equation}
    v(t) = \frac{p^*(t)-(a_M(T-\Delta t)+\alpha)}{\Delta t},
\label{eq:4}
\end{equation}
where the quoted price for both the market sell orders and the limit sell orders with quoted price lower than the market best bid, $b_M$, is given by $p^*(t)=b_M(T-\Delta t)$. Finally, the cancelled sell orders have the velocity of 
\begin{equation}
    v(t) = \frac{(a_M(T-\Delta t)+\alpha)-p^*(t)}{\Delta t}.
\label{eq:5}
\end{equation}

Now, we can define the momentum, $m$, of an order akin the physical momentum of an object, that is, as the product of its size and velocity: $m=s\cdot v$. Thus, to calculate the momentum of submitted market and limit orders we shall use equations (\ref{eq:2}) and (\ref{eq:4}) for velocity, and for cancelled limit orders -- equations (\ref{eq:3}) and (\ref{eq:5}). 

Let $N_\gamma (t)$ denote the total number of limit orders at depth $\gamma$ of the LOB at time $t$. We then calculate the \emph{cumulative sum of net momentum} by aggregating the momentum resulting from all order activities occurred in the active area during a sampling period, $(T-\Delta t,T]$: 
\begin{equation}
\label{eq:active area cumu sum momentum}
    M = \sum_{t \in (T-\Delta t,T]} \sum_{\gamma = b_M-\alpha}^{a_M+\alpha} \sum_{}^{N_\gamma (t)} m.
\end{equation}
 Importantly, our approach models the LOB as a complex system, and emphasizes the vicinity surrounding the bid-ask spread, where order activities exhibit the highest degree of activity. 

\section{Data}
\label{Data}
\noindent
In this section, we describe the cryptocurrency trading data utilized in our experiments, detailing their source and format.

Level 3 order book data were procured from the Coinbase exchange via a websocket feed, encompassing Level 3 trading data for the BTC/USD and LUNA/USD cryptocurrency pairs. As the most granular form of the order book data, Level 3 data includes records of order submissions, cancellations, and matches. In contrast, Level 1 data comprises the Open/High/Low/Close price and volume data for a given period, while Level 2 data incorporates additional information regarding bid and ask order depths.

Orders received at a price equal to or worse than an existing order (typically, the best bid/ask) are executed immediately, generating a 'match' record. Unexecuted or partially executed orders become limit orders in the LOB, creating an 'open' record. These limit orders, if not excuted, will remain on the LOB until they are cancelled, leaving a 'cancelled' record. In our dataset, approximately 98.6\% of open orders are eventually cancelled.

The minimum price precision is 0.01 USD for both BTC/USD and LUNA/USD pairs, and the minimum trading volume unit (order size) is 0.00001 BTC for BTC/USD pair and 0.001 LUNA for LUNA/USD pair. The minimum precision of the recorded timestamp is 0.000001 second (one microsecond). A timestamp is counted when an order activity (or, event) occurs, hence the timestamps of the event-driven records are discrete and not consecutive. For both BTC/USD and LUNA/USD datasets, we grouped the event records by $\Delta t=0.1$ second basis and aggregated all the events occurred in each $\Delta t=0.1$ time period. We also recorded the best bid and best ask prices when the last event of each $\Delta t=0.1$ time period occurs as the best bid/ask prices of the consecutive $\Delta t=0.1$ time period.

The study examined Level 3 trading data for the BTC/USD pair between 20/04/2022 23:00-00:00 and the LUNA/USD pair between 11/05/2022 16:00-20:00, when the LUNA crash occurred. In the LUNA/USD dataset from 11/05/2022 16:00-17:00, a total of 248,796 order submission and cancellation records were identified, with 233,134 occurring within the active area, 8,302 within the passive area, and 7,360 outside. Consequently, about 97\% of order submissions and cancellations transpired within the active area, and 3\% within the passive area.

{In our study, we analysed four distinct datasets across numerous experiments with different objectives. Each contains significant number of records that are large enough to verify the model. The BTC/USD dataset contained over 2,700,000 records, while the LUNA/USD dataset had 800,000 datapoints. In Section \ref{Market_Behaviour}, we used all four datasets to analyse market behaviour. In Sections \ref{Synthetic Spoofing Order}, we used a LUNA/USD dataset, both original and syntethic, to validate spoofing and layering patterns and detect market manipulation. Moreover, 
we applied the Z-score method to original LUNA/USD datasets for comparison with our method. Finally, 
we used both the original and synthetic BTC/USD datasets to compare the performance of our method and the Z-score method.}



\section{Market Behaviour}
\label{Market_Behaviour}
\noindent
In the realm of physical systems, examining the motion of a single particle is infeasible. However, when considering a vast number of particles within a complex system, statistical rules governing their collective behaviour and motion may be discerned. A parallel can be drawn with an order book dynamics, where individual traders place orders based on their supply and demand needs, resulting in irregular market oscillations and noise. Nevertheless, when aggregating all order records and information, emergent collective behaviours become evident. By observing these collective behaviours, insights into market behaviour can be gleaned, potentially unveiling the underlying factors driving the order book and price dynamics.

\begin{figure}[!ht]
  \centering
  \begin{subfigure}[b]{\columnwidth}
    \includegraphics[width=\columnwidth]{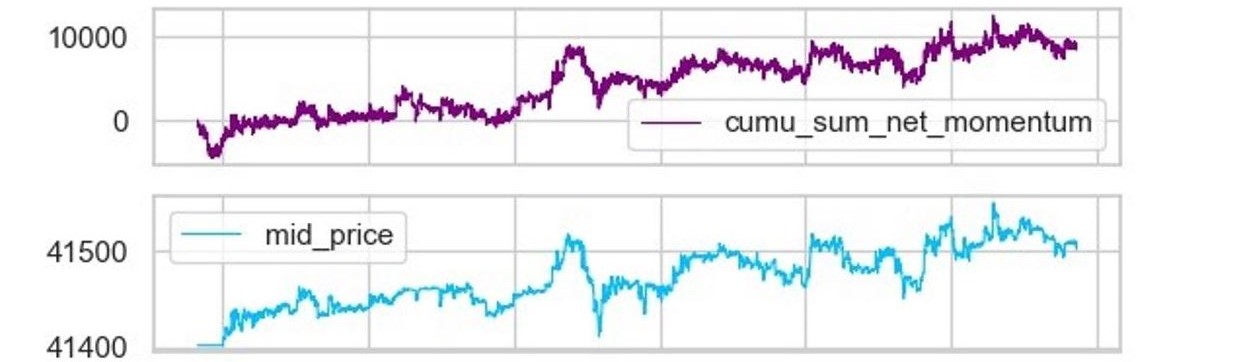} 
    \caption{Limit order and market order unseparated}
    \label{fig:BTC0420_00hr1}
  \end{subfigure}
  \begin{subfigure}[b]{\columnwidth}
    \includegraphics[width=\columnwidth]{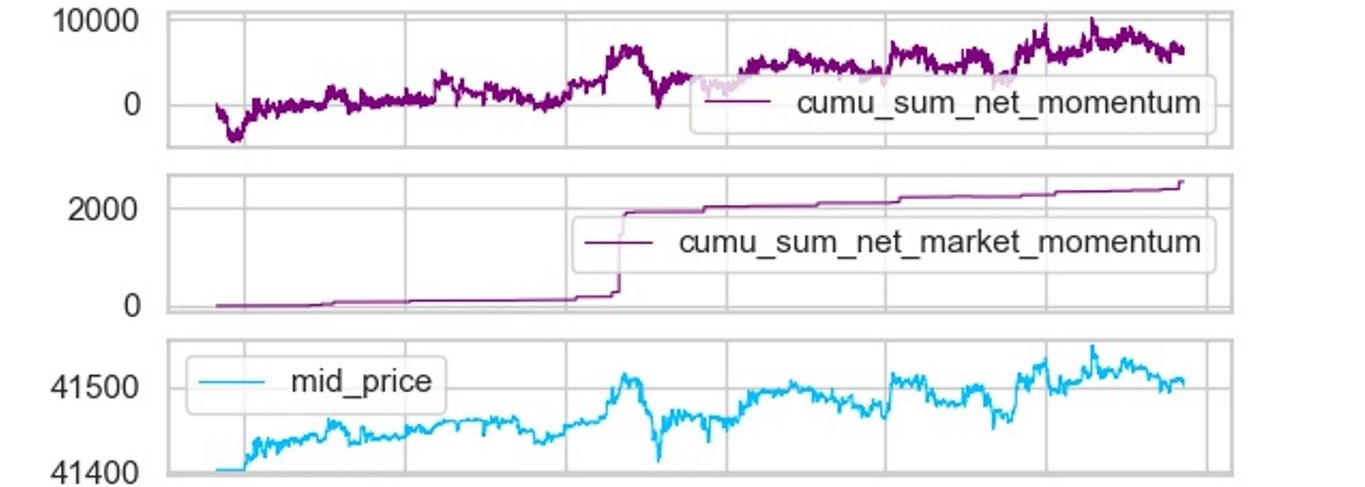}
    \caption{Limit order and market order separated}
    \label{fig:BTC0420_00hr2}
  \end{subfigure}
  \caption{Measures in the active area (depth = 100) for BTC/USD pair during 20/04/2022 23:00-00:00, with limit order and market order unseparated or separated}
\end{figure}

Thus, for the BTC/USD pair during the interval of 20/04/2022 23:00-00:00 and with an active area set to 100, the cumulative sum of net momentum for all orders closely mirrors the midprice movement, as demonstrated in Figure~\ref{fig:BTC0420_00hr1}.

Examining the cumulative sum of net momentum separately for limit orders and market orders yields Figure~\ref{fig:BTC0420_00hr2}, where the first row represents the cumulative sum of net momentum for limit orders and the second row for market orders. We see a noticeable jump in the cumulative sum of net momentum for market orders, which is attributable to the placement of a substantial market buy order 
that does not adhere to the midprice trend.


\begin{figure}[!ht]
  \centering
  \begin{subfigure}[b]{\columnwidth}
    \includegraphics[width=\columnwidth]{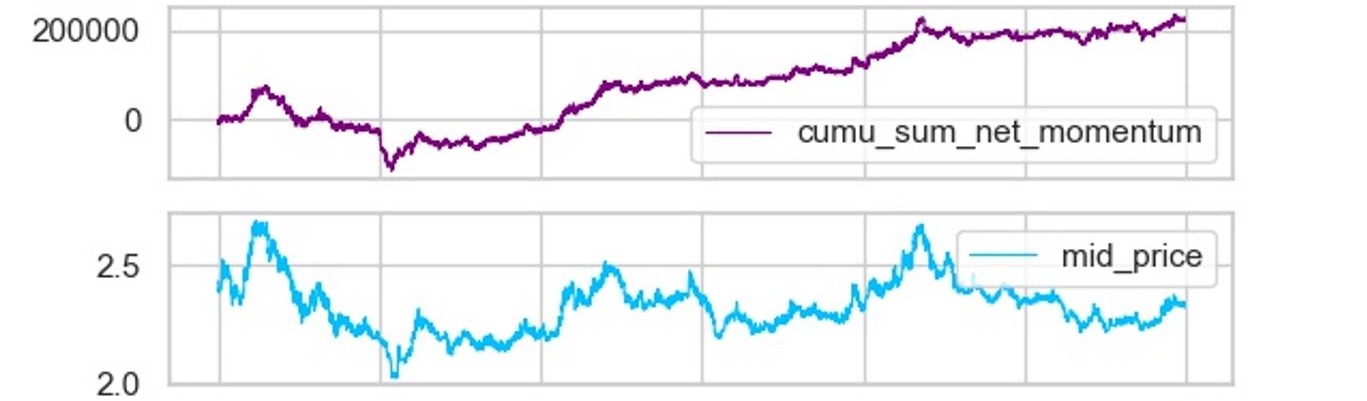} 
    \caption{11/05/2022 16:00-17:00}
    \label{fig:LUNA0511_17hr1}
  \end{subfigure}
  \begin{subfigure}[b]{\columnwidth}
    \includegraphics[width=\columnwidth]{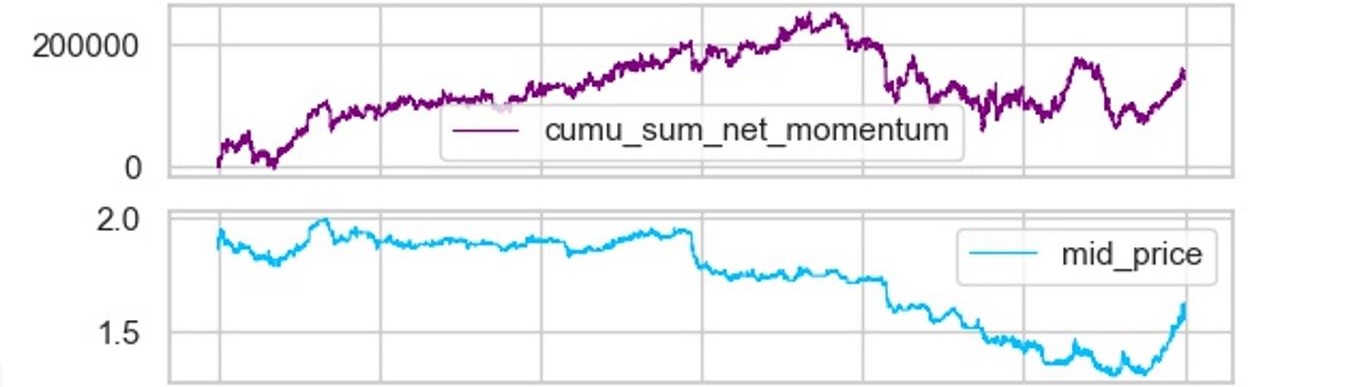}
    \caption{11/05/2022 18:00-19:00}
    \label{fig:LUNA0511_19hr1}
  \end{subfigure}
  \begin{subfigure}[b]{\columnwidth}
    \includegraphics[width=\columnwidth]{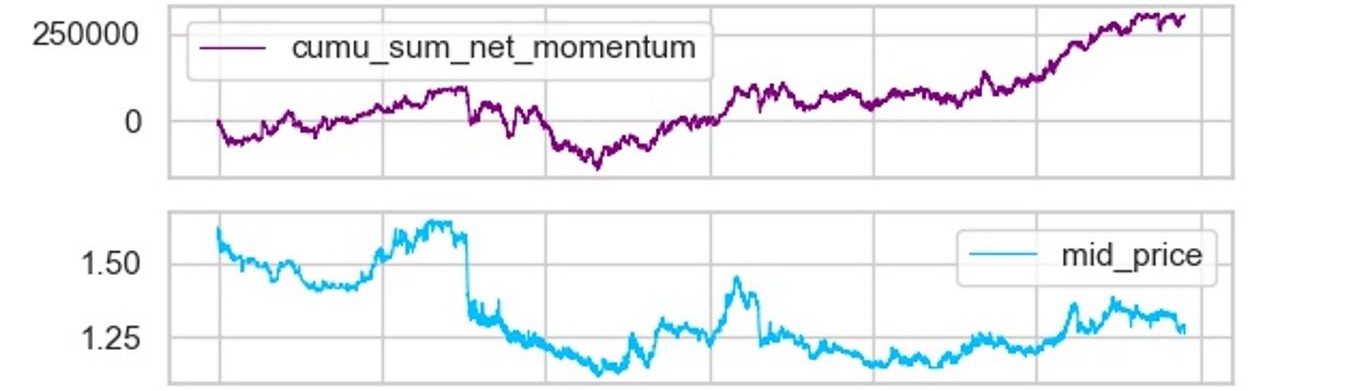}
    \caption{11/05/2022 19:00-20:00}
    \label{fig:LUNA0511_20hr1}
  \end{subfigure}  
  \caption{Measures for LUNA/USD during different periods}
\end{figure}

Similar experiments on the LUNA/USD pair during 11/05/2022 16:00-17:00, 18:00-19:00 and 19:00-20:00 with active area set to 0.5 are illustrated in  Figures~\ref{fig:LUNA0511_17hr1}, ~\ref{fig:LUNA0511_19hr1} and \ref{fig:LUNA0511_20hr1}. 



\begin{figure}[!ht]
  \centering
  \begin{subfigure}[b]{\columnwidth}
    \includegraphics[width=\columnwidth]{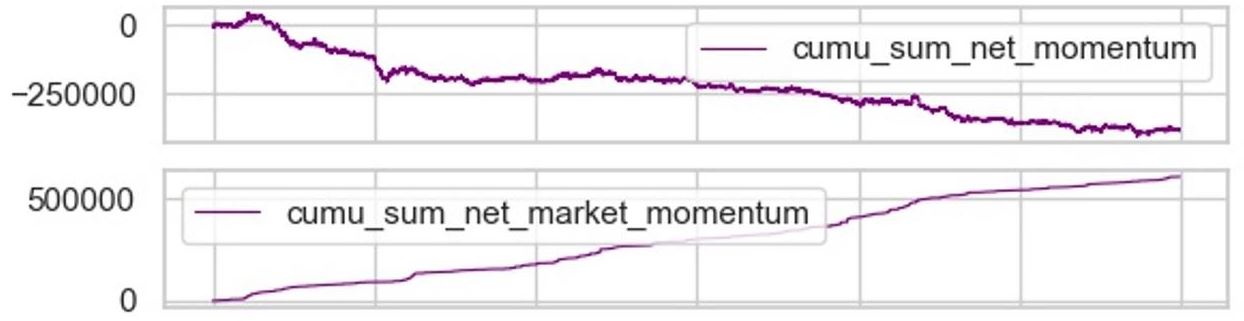} 
    \caption{11/05/2022 16:00-17:00}
    \label{fig:LUNA0511_17hr2}
  \end{subfigure}
  \begin{subfigure}[b]{\columnwidth}
    \includegraphics[width=\columnwidth]{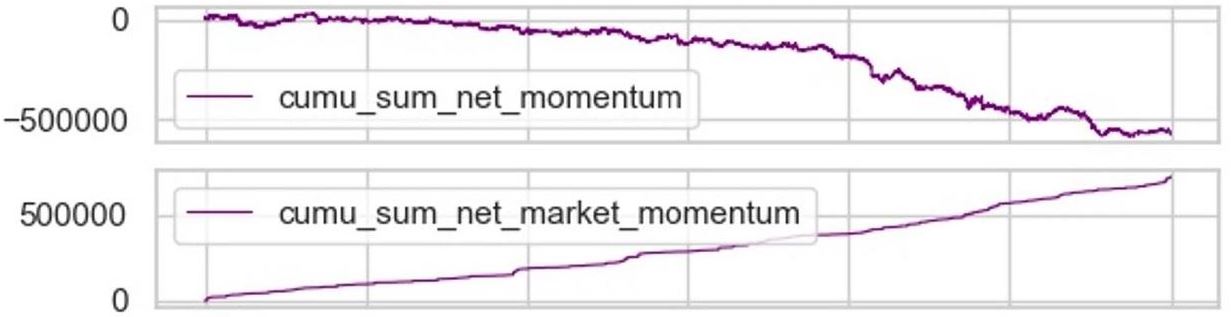}
    \caption{11/05/2022 18:00-19:00}
    \label{fig:LUNA0511_19hr2}
  \end{subfigure}
  \begin{subfigure}[b]{\columnwidth}
    \includegraphics[width=\columnwidth]{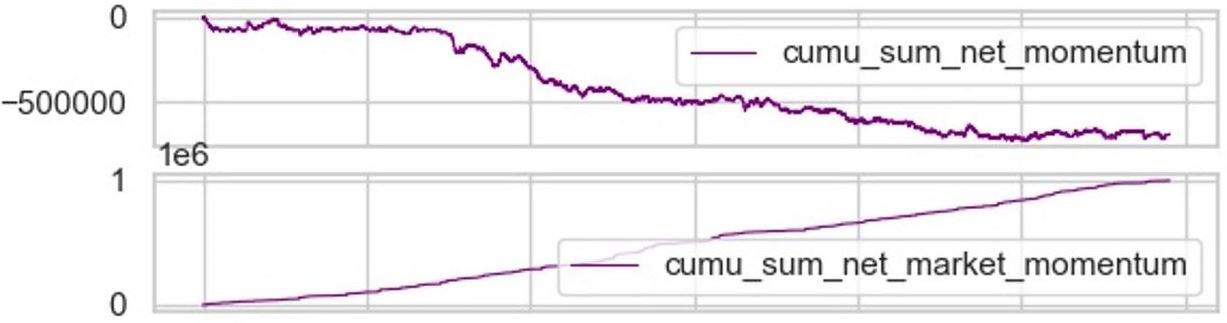}
    \caption{11/05/2022 19:00-20:00}
    \label{fig:LUNA0511_20hr2}
  \end{subfigure}  
  \caption{Measures in the active area (depth = 0.5) for LUNA/USD pair during different periods, with limit order and market order separated}
\end{figure}

The midprice trend is followed albeit somehow more loosely than in the case of BTC/USD (e.g., the midprice downtrends in the middle of Figure \ref{fig:LUNA0511_17hr1} not reflected by the cumulative sum of net momentum). However, upon evaluating the cumulative sum of net momentum separately for limit orders and market orders in Figures \ref{fig:LUNA0511_17hr2}, \ref{fig:LUNA0511_19hr2}, and \ref{fig:LUNA0511_20hr2}, an intriguing market microstructure phenomenon emerges. Market orders consistently exhibit greater momentum than limit orders across all three time periods. Moreover, market orders persistently attempt to drive the price upward (unsuccessfully), while limit orders continuously sell.




\section{Passive Area}
\label{passive area}
In this section, we derive additional insights by turning our attention to the 
\emph{passive area} of the order book, which is the range beyond the active area with the same depth, as depicted in Figure~\ref{fig:foo2}.

Inside the passive area there are only limit orders, whose momentum can be computed by substituting $\alpha$ with $2\alpha$ in equations (\ref{eq:2})--(\ref{eq:5}). The passive cumulative sum of net momentum is then given by
\begin{equation}
\resizebox{.91\linewidth}{!}{$
        \displaystyle
        M^{passive} = \sum_{T \in (t-\Delta t,t]} \sum_{\gamma = b_M-2\alpha}^{b_M-\alpha} \sum_{}^{N_\gamma (T)} m + \sum_{T \in (t-\Delta t,t]} \sum_{\gamma = a_M+\alpha}^{a_M+2\alpha} \sum_{}^{N_\gamma (T)} m
    $}.
\label{eq:passive area cumu sum momentum}
\end{equation}

\begin{figure}[!ht]
  \centering
  \begin{subfigure}[b]{\columnwidth}
    \includegraphics[width=\columnwidth]{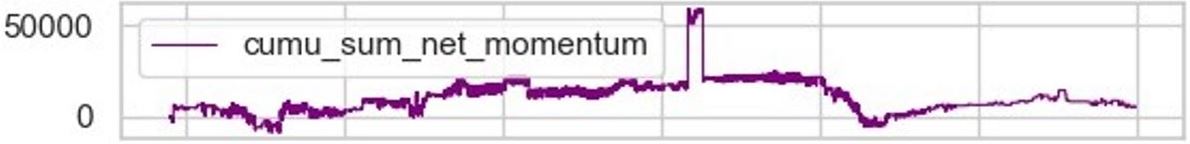} 
    \caption{11/05/2022 16:00-17:00}
    \label{fig:LUNA0511_17hr_passive}
  \end{subfigure}
  \begin{subfigure}[b]{\columnwidth}
    \includegraphics[width=\columnwidth]{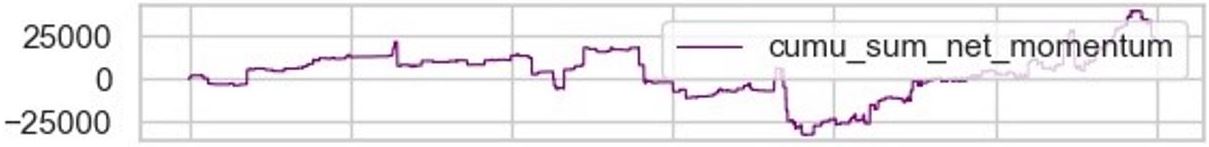}
    \caption{11/05/2022 18:00-19:00}
    \label{fig:LUNA0511_19hr_passive}
  \end{subfigure}
  \begin{subfigure}[b]{\columnwidth}
    \includegraphics[width=\columnwidth]{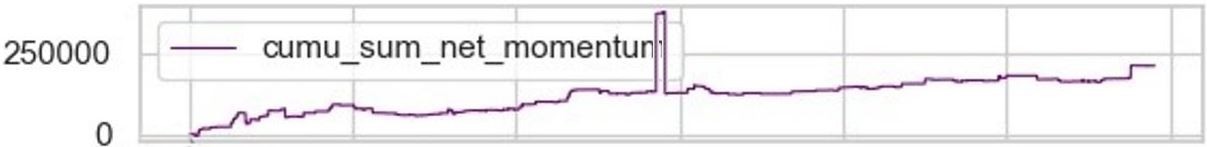}
    \caption{11/05/2022 19:00-20:00}
    \label{fig:LUNA0511_20hr_passive}
  \end{subfigure}  
  \caption{Measures in the passive area for LUNA/USD pair during different periods}
\end{figure}

The measures on the passive area of LUNA/USD pair during 11/05/2022 16:00-17:00, 18:00-19:00, and 19:00-20:00 are shown in Figures~\ref{fig:LUNA0511_17hr_passive},~\ref{fig:LUNA0511_19hr_passive}, and~\ref{fig:LUNA0511_20hr_passive} respectively. Note that the momentum in the passive areas (Figures~\ref{fig:LUNA0511_17hr_passive},~\ref{fig:LUNA0511_19hr_passive},~\ref{fig:LUNA0511_20hr_passive}) 
is much smaller than in the corresponding active areas  (Figures~\ref{fig:LUNA0511_17hr2},~\ref{fig:LUNA0511_19hr2},~\ref{fig:LUNA0511_20hr2}), which indicates lower limit order activity in the passive area vs. the active. More importantly though, we can see the jumps in the momentum that soon bounce back during 11/05/2022 16:00-17:00 and 19:00-20:00 (Figures~\ref{fig:LUNA0511_17hr_passive} and~\ref{fig:LUNA0511_20hr_passive}). It turns out that these jumps indicate the presence of spoofing orders, as discussed in the following Section \ref{Anomaly_Detection}.




The passive area is indeed very useful to detect market manipulations, such as spoofing. As discussed in Section~\ref{Model}, the active area is a range of specific depths in the order book, containing the majority of order submissions and cancellations. The large number of order activities in the active area leads to a significant large value of the momentum. Spoofing events, if any, may then be hidden by the activity of these orders. The smaller order activities and smaller value of momentum in the passive area can facilitate the detection of spoofing by analysing the jumps in the momentum $M^{passive}$ as we will show below. We would also argue that it is less likely for the spoofing orders to be placed inside the active area as they would probably get executed due to the high liquidity around the bid-ask spread. As a result, spoofers would normally tend to place the spoofing orders outside the active area but not very far from the active area for otherwise they would not mislead the other market participants with the order book imbalance. 

{In our model, the identification of active and passive areas is determined through an analysis of market dynamics and is left to be determined by the observer. The separation should distinguish depths of the order book with most of the order activity (active area) from orders that tend to remain inactive (passive area). This distinction is well known in finance (active vs passive traders). To the best of our knowledge there is no known automated algorithm that determines this separation.}

\section{Spoofing Detection}
\label{Spoofing Detection}
\noindent
In this section, we apply our framework to develop an algorithm for detection of spoofing and layering activities in financial markets. 

{To clarify, the physical momentum measure is computed using Equation~\ref{eq:active area cumu sum momentum} in Section~\ref{Model}. The behaviour of market and limit orders can be analysed as described in Section~\ref{Market_Behaviour}. Further computations, as outlined in Equation~\ref{eq:passive area cumu sum momentum} in Section \ref{passive area}, facilitate the detection of market manipulation by identifying anomalies, as described in the Market Manipulation Detection part of Section~\ref{Anomaly_Detection}.}

Spoofing is a manipulative trading practice that involves placing large orders with no intention of executing them, in order to manipulate the volume of visible limit orders and the market microstructure in the order book, to mislead other traders. Spoofing 
was initially regulated as illegal by the 2010 Dodd-Frank Act in the US \cite{US2010} and the 2014 MiFID II Act in the EU \cite{EU2014}. Generally, there are two kinds of spoofing strategies \cite{mark2019spoofing}: (1) traditional spoofing containing a genuine order and a large spoofing order; (2) layered spoofing containing a genuine order and multiple large spoofing orders placed at different prices. 

A spoofer seeks to create a false impression of demand or supply
, inducing price fluctuations that can be exploited for profit. Upon achieving the desired price movement, the spoofer cancels the non-genuine orders before they can be executed. From an academic standpoint, the implications of spoofing are manifolds. First, it disrupts the price discovery process by generating false information, leading to mispriced assets. Second, it undermines the market integrity by creating an unfair trading environment. Finally, spoofing can exacerbate the market volatility and liquidity constraints, especially during periods of heightened uncertainty.

Layering is a more sophisticated form of market manipulation that entails placing multiple non-genuine orders at different price levels. 
The objective is to create an illusion of depth, influencing other market participants to transact at unfavourable prices. Once the target price is reached, the manipulator cancels the layered orders and profits from the induced price movement. Academically, 
layering 
distorts the price discovery process, exacerbates the market volatility, and diminishes liquidity. 

Normally, spoofing and layering happen in markets with relatively poor liquidity, as 
markets with more liquidity would require much larger -- and hence, costly -- spoofing orders. For this reason, by studying the passive area of the order book, our model could easily detect the anomalies in the less volatile LUNA/USD market, while in the highly volatile BTC/USD market there were hardly any spoofing detected.

\label{Synthetic Spoofing Order}
\noindent\rule{0ex}{3ex}{\bf Synthetic Spoofing Order.} 
{It is important to observe that spoofing and layering are covert operations. Consequently, in the original datasets, neither spoofing nor layering events are labeled or identified. To illustrate and identify these patterns and validate our model, we thus created synthetic Level 3 data, as shown in Tables~\ref{table:synthetic data traditional spoofing} and \ref{table:synthetic data layered spoofing}. After the validation on synthetic data, we will be able to detect such patterns in the original data, as discussed in the following parts of Market Manipulation Detection and Comparison with Traditional Techniques. The exact timing of these synthetic data was arbitrarily chosen to mimic the market time dynamics.}

In order to study the nature of spoofing, we inserted synthetic Level 3 order book data into the dataset of LUNA/USD pair during 11/05/2022 18:00-19:00. The synthetic order data of traditional spoofing listed in Table~\ref{table:synthetic data traditional spoofing} were inserted to timestamps at 11/05/2022 18:36.13.59 and 18:38.16.02. Similarly, the synthetic order data of layered spoofing listed in Table~\ref{table:synthetic data layered spoofing} were inserted to timestamps between 11/05/2022 18:36.13.59 and 18:38.16.02.

\begin{figure}[!ht]
  \centering
  \begin{subfigure}[b]{\columnwidth}
    \includegraphics[width=\columnwidth]{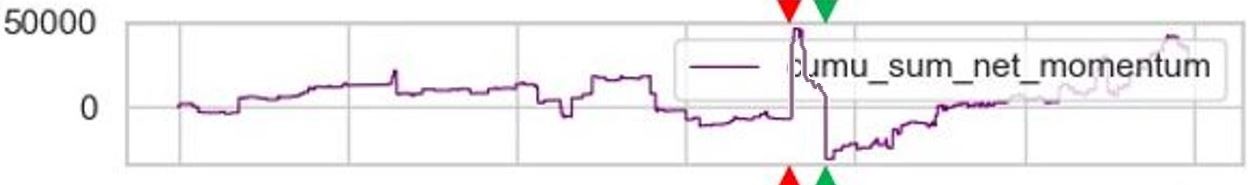} 
    \caption{Synthetic data of traditional spoofing}
    \label{fig:LUNA0511_19hr_traditional_spoofing}
  \end{subfigure}
  \begin{subfigure}[b]{\columnwidth}
    \includegraphics[width=\columnwidth]{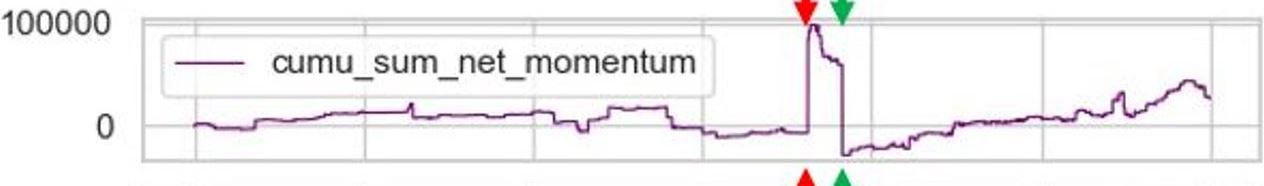}
    \caption{Synthetic data of layered spoofing}
    \label{fig:LUNA0511_19hr_layered_spoofing}
  \end{subfigure}
  \caption{Measures in the passive area for LUNA/USD pair during 11/05/2022 18:00-19:00, with synthetic data of traditional spoofing and layered spoofing}
\end{figure}

The best bid and ask prices were 1.74 and 1.75, respectively, between 11/05/2022 18:36.13.59 and 18:38.16.02. As the price of synthetic spoofing orders is located outside of the active area (1.24 -- 2.25), they will not have any impact on the measures in this range. The measures in the passive area on the datasets enriched with synthetic traditional and layered spoofing orders are shown in Figures~\ref{fig:LUNA0511_19hr_traditional_spoofing} and~\ref{fig:LUNA0511_19hr_layered_spoofing}, respectively, where red arrows indicate the time of placing the orders, and green arrows -- of their cancellations.

\begin{table}
\begin{small}
\begin{center}
    \caption{The synthetic data}    
    \begin{subtable}[h]{0.45\textwidth}
        \centering
        \begin{tabular}{lrrrr}
        \hline
        \rule{0pt}{6pt}
        Timestamp & Price & Order Type	& Side & Size  \\ 
        \hline
        \\ [-6pt]
        18:36.13.59	& 1.14 & limit & buy & 100000\\
        18:38.16.02	& 1.14 & cancel & buy & 100000 \\
        \hline
       \end{tabular}
       \caption{The synthetic data of traditional spoofing}
       \label{table:synthetic data traditional spoofing}
    \end{subtable}
    \hfill
    \begin{subtable}[h]{0.45\textwidth}
        \centering
        \begin{tabular}{lrrrr}
        \hline
        \rule{0pt}{6pt}
        Timestamp & Price & Order Type	& Side & Size  \\ 
        \hline
        \\ [-6pt]
        18:36.13.59	& 1.20 & limit & buy & 50000 \\
        18:36.14.44	& 1.12 & limit & buy & 50000 \\
        18:36.15.38	& 1.04 & limit & buy & 50000 \\
        18:36.16.23	& 0.96 & limit & buy & 50000 \\
        18:38.16.02	& 1.20 & cancel & buy & 50000 \\
        18:38.16.02	& 1.12 & cancel & buy & 50000 \\
        18:38.16.02	& 1.04 & cancel & buy & 50000 \\
        18:38.16.02	& 0.96 & cancel & buy & 50000 \\
        \hline
        \end{tabular}
        \caption{The synthetic data of layered spoofing}
        \label{table:synthetic data layered spoofing}
     \end{subtable}
\end{center}
\end{small}
\end{table}

Comparing Figures~\ref{fig:LUNA0511_19hr_traditional_spoofing} and~\ref{fig:LUNA0511_19hr_layered_spoofing} to Figure~\ref{fig:LUNA0511_19hr_passive}, we see how spoofing events can be easily visualised by rapid jumps of the momentum. 
Furthermore, our method significantly reduces the complexity of the mathematical model and computational techniques to recognise and detect spoofing, as discussed in the following Section \ref{Anomaly_Detection}. We observe that the layered spoofing has the same structure and pattern as the traditional spoofing, as the jumps of momentum displayed for the layered spoofing are the same as that for the traditional spoofing, except for their fluctuating amplitude. This is because the submission and cancellation time of synthetic traditional spoofing and layered spoofing orders were about the same.



\label{Anomaly_Detection}
\smallskip \noindent {\bf Market Manipulation Detection.}
After computing the net momentum for all order activities (including submissions and cancellations), it is possible to compute the statistical mean and standard variance of the time series of the cumulative net momentum at any sampling frequency: $\texttt{deviation} = \frac{\texttt{net momentum} - \texttt{mean}}{\texttt{standard variance}}$. 
Thus, spoofing orders can be simply detected and located using equation \ref{eq:passive area cumu sum momentum}, by inspecting the largest anomalous deviation value in any time period. For example, among the 36,000 sampling intervals in the LUNA/USD dataset during 11/05/2022 19:00-20:00, the largest ten deviation values (rounded to two decimal places) were: 
    [126.12 120.76 17.72 7.09 7.02 7.02 6.80 6.80 6.80 6.61].

\begin{figure}[!ht]
  \centering
  \begin{subfigure}[b]{\columnwidth}
    \includegraphics[width=\columnwidth]{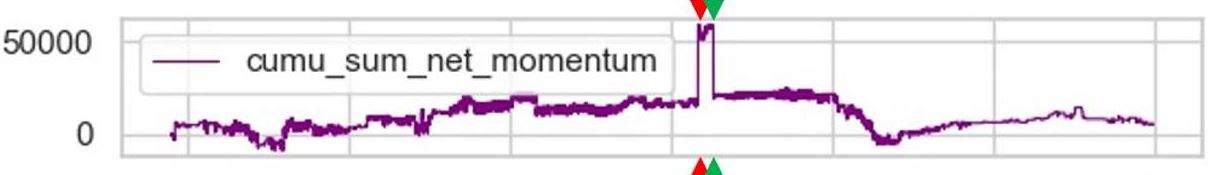} 
    \caption{11/05/2022 16:00-17:00, with traditional spoofing detected}
    \label{fig:LUNA0511_17hr_passive_detected}
  \end{subfigure}
  \begin{subfigure}[b]{\columnwidth}
    \includegraphics[width=\columnwidth]{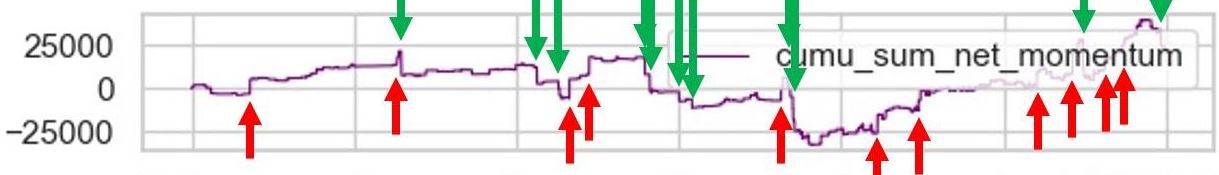}
    \caption{11/05/2022 18:00-19:00, with layered spoofing detected}
    \label{fig:LUNA0511_19hr_passive_detected}
  \end{subfigure}
  \begin{subfigure}[b]{\columnwidth}
    \includegraphics[width=\columnwidth]{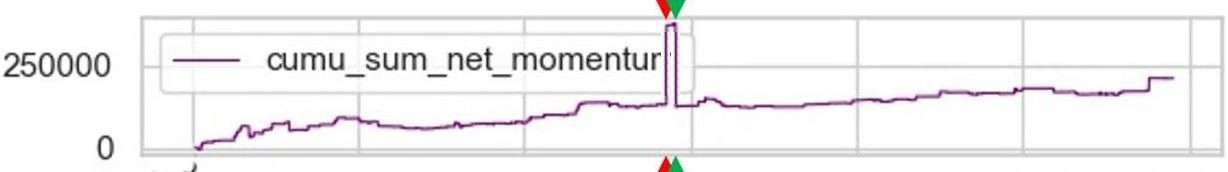}
    \caption{11/05/2022 19:00-20:00, with traditional spoofing detected}
    \label{fig:LUNA0511_20hr_passive_detected}
  \end{subfigure}  
  \caption{Measures in the passive area for LUNA/USD pair during different period, with market manipulation activity detected}
\end{figure}

The two largest deviation values detected as anomaly correspond to the timestamps 19:28:34.77 and 19:29:07.40, which are exactly where the jump and bounce back happened in Figure~\ref{fig:LUNA0511_20hr_passive_detected}.



It is feasible to trace the corresponding anomaly time intervals in the original Level 3 order book data, and find the order records related to spoofing. They are listed in Table \ref{table:spoofing_detected_17hr}.


\begin{table}
\begin{small}
\begin{center}
    \caption{Spoofing detected for LUNA/USD}    
    \begin{subtable}[h]{0.45\textwidth}
        \centering
        \begin{tabular}{lrrrr}
        \hline
        \rule{0pt}{6pt}
        Timestamp & Price & Order Type	& Side & Size  \\ 
        \hline
        \\ [-6pt]
        16:31:39.50	& 1.62 & limit & buy & 111939.762\\
        16:32:37.06	& 1.62 & cancel & buy & 111939.762\\
        \hline
       \end{tabular}
       \caption{11/05/2022 16:00-17:00}
       \label{table:spoofing_detected_17hr}
    \end{subtable}
    \hfill
    \begin{subtable}[h]{0.45\textwidth}
        \centering
        \begin{tabular}{lrrrr}
        \hline
        \rule{0pt}{6pt}
        Timestamp & Price & Order Type	& Side & Size  \\ 
        \hline
        \\ [-6pt]
        18:03:32.66	& 1.10 & limit & buy & 29000\\ 
        18:12:52.87	& 1.36 & cancel & buy & 29000\\
        18:23:14.53	& 1.25 & limit & buy & 29000\\
        18:24:27.68	& 1.24 & limit & buy & 29000\\ 
        18:27:53.69	& 1.25 & cancel & buy & 29000\\
        18:28:11.94	& 1.24 & cancel & buy & 29000\\
        18:42:14.77	& 1.06 & limit & buy & 20000\\       
        18:54:15.66	& 0.80 & limit & buy & 29000\\
        18:54:54.42	& 0.80 & cancel & buy & 29000\\
        18:59:42.93	& 1.01 & cancel & buy & 20000\\ 
        \hline
        \end{tabular}
        \caption{11/05/2022 18:00-19:00}
        \label{table:spoofing_detected_19hr}
     \end{subtable}
    \hfill
    \begin{subtable}[h]{0.45\textwidth}
        \centering
        \begin{tabular}{lrrrr}
        \hline
        \rule{0pt}{6pt}
        Timestamp & Price & Order Type	& Side & Size  \\ 
        \hline
        \\ [-6pt]
        19:28:34.77	& 0.73 & limit & buy & 535665.177\\
        19:29:07.48	& 0.73 & cancel & buy & 535665.177\\
        \hline
        \end{tabular}
        \caption{11/05/2022 19:00-20:00}
        \label{table:spoofing_detected_20hr}
     \end{subtable}
\end{center}
\end{small}
\end{table}

Interestingly, the cumulative sum of net momentum in the passive area for LUNA/USD pair in Figure~\ref{fig:LUNA0511_19hr_passive_detected} is relatively less 'smooth' or 'continuous' than that in Figure~\ref{fig:LUNA0511_17hr_passive_detected} and Figure~\ref{fig:LUNA0511_20hr_passive_detected}. This measure visually displays many irregular small jumps during 18:00-19:00, although the magnitudes are less than the large jumps during 16:00-17:00 and 19:00-20:00. The irregular jumps during 18:00-19:00 are depicted in Figure~\ref{fig:LUNA0511_19hr_passive_detected} where red arrows indicate increasing jumps of the momentum, and green arrows -- decreasing jumps.

The anomaly detection algorithm signals the largest ten deviation value (rounded to two decimal places) listed as follows:
    [49.51 48.48 43.31 40.21 36.08 35.05 33.41 33.01 32.98 29.88]. 
The traced corresponding order records are listed in Table \ref{table:spoofing_detected_19hr}, where most of the detected suspicious order records have the same order size of 29,000. As a result, it is reasonable to infer that these detected spoofing orders are all part of the layered spoofing. Furthermore, by examining all order records with exact order size of 29,000, we found 124 records in total during 11/05/2022 18:00-19:00. Almost all these records were limit buy orders that got cancelled soon, and most of them were placed in the passive area. 


\label{Comparison1}
\smallskip \noindent {\bf Comparison with Traditional Techniques.}
We compare our method with the traditional anomaly detection technique, Z-score.

Among the 36,000 sampling intervals in the LUNA/USD dataset during 11/05/2022 16:00-17:00, the largest ten deviation values (rounded to two decimal places) were: [43.31 40.89 8.78 8.75 8.50 7.92 7.13 6.51 6.47 5.73]. The two largest deviation values detected as anomaly correspond to the timestamps 16:31:39.50 and 16:32:37.00, which are exactly where the jump and bounce back happened in Figure~\ref{fig:LUNA0511_17hr_passive_detected}. 
It is easy to trace the corresponding anomaly time intervals in the raw Level 3 order book data, and find the order records related to spoofing. They are listed in Table \ref{table:spoofing_detected_17hr}.

Note that at these two timestamps, the best bid and ask prices of LUNA/USD pair were 2.26 and 2.28 respectively. These spoofing orders were quoted at 1.62 inside the passive area and have significant influence to the market.

The Z-score methodology is a prominent statistical approach, it is employed to detect outliers within a dataset by quantifying the deviation of individual data points from the mean value in units of standard deviations. Often referred to as the standard score, the Z-score is given by $Z = \frac{X-\mu}{\sigma}$ where $X$ is the data point being evaluated, $\mu$ is the mean of the dataset, and $\sigma$ is the standard deviation of the dataset. Using the Z-score technique on the same LUNA/USD dataset during 11/05/2022 16:00-17:00, there are 768,270 samples of order records, and the largest ten Z-scores among them: [378.01 9.23 1.68 1.68 1.68 1.68 1.68 1.68 1.68 1.68]. After inspecting the two samples with the largest Z-score, the found related order records are listed in Table \ref{table:z_score_LUNA}.

\begin{table}
\begin{small}
\begin{center}
    \begin{tabular}{lrrrr}
        \hline
        \rule{0pt}{6pt}
        Timestamp & Price & Order Type	& Side & Size  \\ 
        \hline
        \\ [-6pt]
        16:00:42.80	& 0.01 & limit & buy & 356402.418\\
        16:10:32.77	& 0.11 & limit & buy & 6374711.154\\
        \hline
    \end{tabular}
    \caption{Order records detected by Z-score technique on LUNA/USD pair during 11/05/2022 16:00-17:00}
    \label{table:z_score_LUNA}
\end{center}
\end{small}
\end{table}

\begin{figure}[!ht]
\begin{center}
\includegraphics[width=0.8\columnwidth]{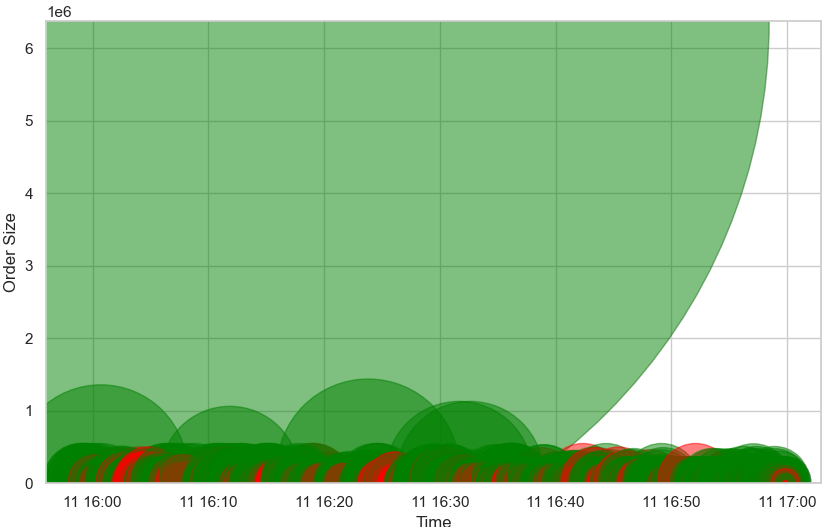}
\end{center}
\caption{Visualisation of the order activities on LUNA/USD pair during 11/05/2022 16:00-17:00}
\label{fig:visual_LUNA}
\end{figure}

Figure~\ref{fig:visual_LUNA} is another way to visualise the order activities during the same period using the raw Level 3 order book data, which contains 768,271 order records in total. Each circle represents an order record, either submission or cancellation, while the radius of each circle is equal to its corresponding order size. Green circles are buy orders and red ones are sell orders. We see that the spoofing orders mentioned above in Table \ref{table:spoofing_detected_17hr} are shadowed by the other orders with larger size and infeasible to recognise directly.

In contrast, the two orders identified by the Z-score technique exhibited substantial size, with the corresponding best bid and ask prices at 2.43 \& 2.46 and 2.10 \& 2.12, respectively. Evidently, these quotes fell outside the pre-established passive area and were too distant from the bid-ask spread to exert any influence on the market. Furthermore, these orders were not accompanied by subsequent cancellations, rendering them improbable candidates for spoofing. 

\label{Comparison2}
\smallskip \noindent {\bf Experiments on the Bitcoin Trading Market.}
Below, we compare the physics-based momentum and traditional methods on the BTC/USD cryptocurrency pair. 
\begin{figure}[!ht]
\begin{center}
\includegraphics[width=0.8\columnwidth]{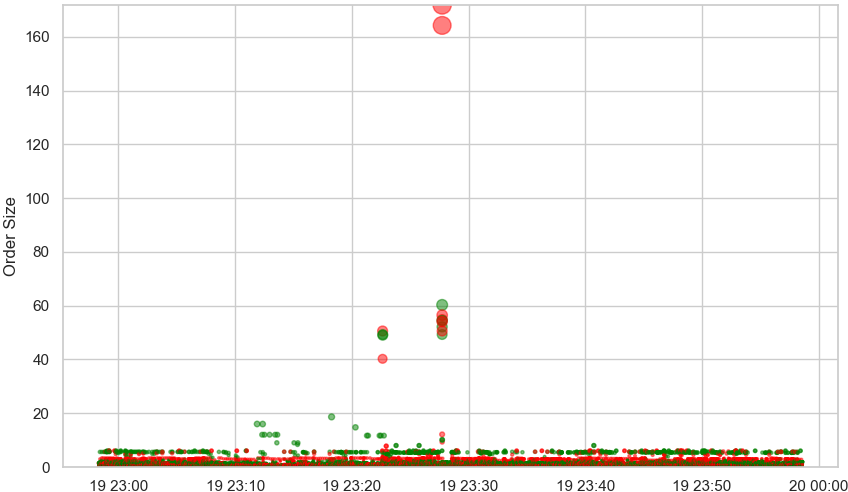}
\end{center}
\caption{Visualisation of the order activities on BTC/USD pair during 19/04/2022 23:00-00:00}
\label{fig:visual_BTC}
\end{figure}

A visualisation is similar to that in Figure~\ref{fig:visual_BTC}, except the order activities during 19/04/2022 23:00-00:00 shown in
Figure~\ref{fig:visual_BTC}, which contains 1,348,156 order records in total. The reason for why the circles in this figure are much smaller than those in Figure~\ref{fig:visual_LUNA} is that the nominal price of Bitcoin (around 40000 USD) is significantly larger than that of LUNA (around 1 USD), which results in the order size with values of different scales. 
{Note that Figures \ref{fig:visual_LUNA} and \ref{fig:visual_BTC} were included to visualise the order size of each limit order over time. These figures illustrate that the detection of spoofing cannot solely rely on order size and underscore the efficacy of our method.}

\begin{figure}[!ht]
\begin{center}
\includegraphics[width=\columnwidth]{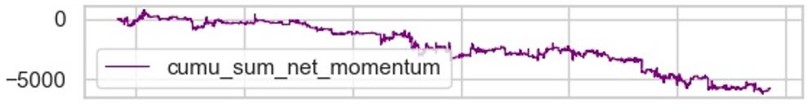}
\end{center}
\caption{Measures in the passive area for BTC/USD pair during 19/04/2022 23:00-00:00}
\label{fig:BTC_passive}
\end{figure}

Figure~\ref{fig:BTC0420_00hr2} shows the cumulative sum of momentum within the active area, 
and Figure~\ref{fig:BTC_passive} of the measure inside the passive area.
In this dataset, the largest ten deviation values (rounded to two decimal places) on the momentum inside the passive area were: [26.44 26.24 25.13 23.05 22.766 22.62 20.68 19.57 19.38 19.04]. There were no  suspicious orders 
according to Figure~\ref{fig:BTC_passive}, which was justified by examining the order records relative to the above data. 

However, with applying the Z-score technique on the raw Level 3 order book data of all the quoted order size, the largest ten Z-scores (rounded to two decimal places) among them were: [287.55 287.55 274.83 274.83 100.68 94.28 91.40 91.27 90.87 90.27]. The corresponding order records are listed in Table \ref{table:z_score_BTC}.
\begin{table}
\begin{small}
\begin{center}
    \begin{tabular}{lrrrr}
        \hline
        \rule{0pt}{6pt}
        Timestamp & Price & Order Type	& Side & Size  \\ 
        \hline
        \\ [-6pt]
        23:22:37.77	& 41688.74 & limit & sell & 50.589\\
        23:27:43.83	& 41259.66 & limit & buy & 52.116\\
        23:27:43.84	& 40985.33 & cancel & buy & 54.433\\
        23:27:43.84	& 41729.97 & cancel & sell & 54.063\\
        23:27:43.85	& 41839.44 & limit & sell & 54.745\\
        23:27:43.85	& 41094.08 & limit & buy & 60.284\\
        23:27:43.87	& 42288.36 & cancel & sell & 171.821\\
        23:27:43.89	& 42398.82 & limit & sell & 164.23\\
        23:27:43.95	& 41688.74 & cancel & sell & 50.589\\
        23:27:43.97	& 41673.86 & limit & sell & 56.484\\
        \hline
    \end{tabular}
    \caption{Order records detected by Z-score technique on BTC/USD pair during 19/04/2022 23:00-00:00}
    \label{table:z_score_BTC}
\end{center}
\end{small}
\end{table}

Note that although the limit sell order quoted on 41688.74 place at 23:22:37.77 was cancelled at 23:27:43.95, it is not likely to be spoofing as its quoted price was too far (more than 200 USD) from the midprice which was 41466.87 during the lifetime of this order.

In order to further compare these two methods, we inserted synthetic order book data of traditional spoofing into the BTC/USD dataset. The synthetic order records were inserted as Table~\ref{table:synthetic_data_BTC} with the synthetic spoofing orders quoted inside the passive area.

\begin{table}
\begin{small}
\begin{center}
    \begin{tabular}{lrrrr}
        \hline
        \rule{0pt}{6pt}
        Timestamp & Price & Order Type	& Side & Size  \\ 
        \hline
        \\ [-6pt]
        23:23.22.81	& 41334.00 & limit & buy & 40\\
        23:24.42.68	& 41334.00 & cancel & buy & 40 \\
        \hline
    \end{tabular}
    \caption{The synthetic data of traditional spoofing for BTC/USD}
    \label{table:synthetic_data_BTC}
\end{center}
\end{small}
\end{table}

\begin{figure}[!ht]
\begin{center}
\includegraphics[width=\columnwidth]{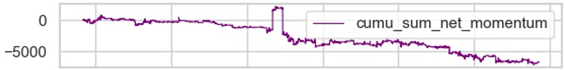}
\end{center}
\caption{Measures in the passive area for BTC/USD pair during 19/04/2023 23:00-00:00, with synthetic data of traditional spoofing}
\label{fig:BTC_passive_syn}
\end{figure}

We observe an apparent abnormal jump (see Figure~\ref{fig:BTC_passive_syn}) compared to the original dataset shown in Figure~\ref{fig:BTC_passive}.

In this synthetic dataset, the largest ten deviation values (rounded to two decimal places) of the momentum inside the passive area were: [60.75 52.23 23.84 23.66 22.66 20.78 20.52 20.39 18.65 17.65]. These values are noticeably different from those of the original dataset, and by examining the anomaly data sample we could effortlessly locate the spoofing orders listed in Table~\ref{table:synthetic_data_BTC}.

In contrast, the Z-score (rounded to two decimal places) of this synthetic dataset remains almost the same compared to that of the original dataset: [286.12 286.12 273.46 273.46 100.17 93.81 90.94 90.81 90.42 89.82]. We thus conclude that for the BTC/USD pair, the physics-based momentum method outperforms the traditional Z-score-based method in analysing the Level 3 order book data.

\label{Comparative analysis}
\smallskip \noindent {\bf Comparative Analysis.} 
{Regarding the accuracy and false positive rate of our model, our method has shown positive results, especially when compared to the Z-score method. Notably, our method successfully detected spoofing events in the original dataset of LUNA trading market as listed in Table \ref{table:spoofing_detected_17hr}, whereas the Z-score method, using the same dataset, resulted in the findings presented in Table \ref{table:z_score_LUNA} that are normal orders instead of spoofing. This is true also for the synthetic dataset of Bitcoin trading market, comparing Table \ref{table:synthetic_data_BTC} with Table \ref{table:z_score_BTC}. The Z-score method not only failed to identify spoofing but also incorrectly flagged large limit orders as spoofing.}

{This comparative analysis substantiates the infeasibility of directly detecting spoofing from raw Level 3 order book data as the spoofing orders are likely to be overshadowed by larger-volume-orders without purpose of spoofing. Consequently, these results demonstrate that the proposed physics-based momentum methodology offers a valid means for modeling Level 3 order book data, thereby enabling the extraction and analysis of pertinent information regarding the order book and market dynamics.}



\section{Conclusion}
\label{Conclusion}

The analysis conducted in this study employed Level 3 order book data, as opposed to time-based snapshot data, enabling a more granular examination of the market microstructure without sacrificing temporal information. 
Orders were modelled as physical particles, with the active area and passive area concepts employed to compute each order's momentum and assess its impact on the order book. Consequently, a systematic representation of the order book state was established, utilizing the extensive Level 3 order book data both feasibly and efficiently. The market behaviour analysis revealed contrasting actions of limit and market orders during the LUNA cryptocurrency flash crash. These visualizations also provide insights into the market patterns, offering valuable indicators for detecting market manipulation. Messages are depicted in a distinctive and intuitive manner, facilitating the retrospective identification of malicious orders. The presence of spoofing and layering was uncovered in the LUNA cryptocurrency trading market.

{From the AI perspective, our model offers a novel method for bridging statistical physics and financial market analysis. This has the potential to enhance existing AI models for financial markets, particularly those utilising machine learning techniques. Specifically, our work proposes an innovative way to process and label Level 3 order book data (potential pernicious order or not), allowing for a more in-depth study of market microstructure.}

In conclusion, the approach outlined in this paper may serve as a foundation for future research exploring 
additional physical concepts 
and examining market manipulation across other asset classes, including equities and futures.




    
\bibliography{reference}

\begin{thebibliography}{}

\bibitem [\protect \citeauthoryear {%
Cartea%
, Jaimungal%
\BCBL {}\ \BBA {} Wang%
}{%
Cartea%
\ \protect \BOthers {.}}{%
{\protect \APACyear {2020}}%
}]{%
cartea2020spoofing}
\APACinsertmetastar {%
cartea2020spoofing}%
\begin{APACrefauthors}%
Cartea, {\'A}.%
, Jaimungal, S.%
\BCBL {}\ \BBA {} Wang, Y.%
\end{APACrefauthors}%
\unskip\
\newblock
\APACrefYearMonthDay{2020}{}{}.
\newblock
{\BBOQ}\APACrefatitle {Spoofing and price manipulation in order-driven markets}
  {Spoofing and price manipulation in order-driven markets}.{\BBCQ}
\newblock
\APACjournalVolNumPages{Applied Mathematical Finance}{27}{1-2}{67--98}.
\PrintBackRefs{\CurrentBib}

\bibitem [\protect \citeauthoryear {%
{European Commission Banking and Finance}%
}{%
{European Commission Banking and Finance}%
}{%
{\protect \APACyear {2014}}%
}]{%
EU2014}
\APACinsertmetastar {%
EU2014}%
\begin{APACrefauthors}%
{European Commission Banking and Finance}.%
\end{APACrefauthors}%
\unskip\
\newblock
\APACrefYearMonthDay{2014}{}{}.
\newblock
\APACrefbtitle {Markets in Financial Instruments Directive 2014/65/EU (MiFID
  2).} {Markets in financial instruments directive 2014/65/eu (mifid 2).}
\newblock
\APAChowpublished
  {\url{https://eur-lex.europa.eu/legal-content/EN/TXT/PDF/?uri=CELEX:32014L0065}}.
\PrintBackRefs{\CurrentBib}

\bibitem [\protect \citeauthoryear {%
Farmer%
, Patelli%
\BCBL {}\ \BBA {} Zovko%
}{%
Farmer%
\ \protect \BOthers {.}}{%
{\protect \APACyear {2005}}%
}]{%
farmer2005predictive}
\APACinsertmetastar {%
farmer2005predictive}%
\begin{APACrefauthors}%
Farmer, J\BPBI D.%
, Patelli, P.%
\BCBL {}\ \BBA {} Zovko, I\BPBI I.%
\end{APACrefauthors}%
\unskip\
\newblock
\APACrefYearMonthDay{2005}{}{}.
\newblock
{\BBOQ}\APACrefatitle {The predictive power of zero intelligence in financial
  markets} {The predictive power of zero intelligence in financial
  markets}.{\BBCQ}
\newblock
\APACjournalVolNumPages{Proceedings of the National Academy of
  Sciences}{102}{6}{2254--2259}.
\PrintBackRefs{\CurrentBib}

\bibitem [\protect \citeauthoryear {%
G{\"u}nnemann%
, G{\"u}nnemann%
\BCBL {}\ \BBA {} Faloutsos%
}{%
G{\"u}nnemann%
\ \protect \BOthers {.}}{%
{\protect \APACyear {2014}}%
}]{%
gunnemann2014robust}
\APACinsertmetastar {%
gunnemann2014robust}%
\begin{APACrefauthors}%
G{\"u}nnemann, N.%
, G{\"u}nnemann, S.%
\BCBL {}\ \BBA {} Faloutsos, C.%
\end{APACrefauthors}%
\unskip\
\newblock
\APACrefYearMonthDay{2014}{}{}.
\newblock
{\BBOQ}\APACrefatitle {Robust multivariate autoregression for anomaly detection
  in dynamic product ratings} {Robust multivariate autoregression for anomaly
  detection in dynamic product ratings}.{\BBCQ}
\newblock
\BIn{} \APACrefbtitle {Proceedings of the 23rd international conference on
  World wide web} {Proceedings of the 23rd international conference on world
  wide web}\ (\BPGS\ 361--372).
\PrintBackRefs{\CurrentBib}

\bibitem [\protect \citeauthoryear {%
{House of Representatives}%
}{%
{House of Representatives}%
}{%
{\protect \APACyear {2010}}%
}]{%
US2010}
\APACinsertmetastar {%
US2010}%
\begin{APACrefauthors}%
{House of Representatives}.%
\end{APACrefauthors}%
\unskip\
\newblock
\APACrefYearMonthDay{2010}{}{}.
\newblock
\APACrefbtitle {Dodd‐Frank Wall Street Reform and Consumer Protection Act:
  Conference Report (to Accompany H.R. 4173).} {Dodd‐frank wall street reform
  and consumer protection act: Conference report (to accompany h.r. 4173).}
\newblock
\APAChowpublished
  {\url{https://www.congress.gov/111/plaws/publ203/PLAW-111publ203.pdf}}.
\PrintBackRefs{\CurrentBib}

\bibitem [\protect \citeauthoryear {%
Leangarun%
, Tangamchit%
\BCBL {}\ \BBA {} Thajchayapong%
}{%
Leangarun%
\ \protect \BOthers {.}}{%
{\protect \APACyear {2016}}%
}]{%
leangarun2016stock}
\APACinsertmetastar {%
leangarun2016stock}%
\begin{APACrefauthors}%
Leangarun, T.%
, Tangamchit, P.%
\BCBL {}\ \BBA {} Thajchayapong, S.%
\end{APACrefauthors}%
\unskip\
\newblock
\APACrefYearMonthDay{2016}{}{}.
\newblock
{\BBOQ}\APACrefatitle {Stock price manipulation detection based on mathematical
  models} {Stock price manipulation detection based on mathematical
  models}.{\BBCQ}
\newblock
\APACjournalVolNumPages{International journal of trade, economics and
  finance}{7}{3}{81--88}.
\PrintBackRefs{\CurrentBib}

\bibitem [\protect \citeauthoryear {%
Lillo%
\ \BBA {} Doyne~Farmer%
}{%
Lillo%
\ \BBA {} Doyne~Farmer%
}{%
{\protect \APACyear {2005}}%
}]{%
lillo2005key}
\APACinsertmetastar {%
lillo2005key}%
\begin{APACrefauthors}%
Lillo, F.%
\BCBT {}\ \BBA {} Doyne~Farmer, J.%
\end{APACrefauthors}%
\unskip\
\newblock
\APACrefYearMonthDay{2005}{}{}.
\newblock
{\BBOQ}\APACrefatitle {The key role of liquidity fluctuations in determining
  large price changes} {The key role of liquidity fluctuations in determining
  large price changes}.{\BBCQ}
\newblock
\APACjournalVolNumPages{Fluctuation and Noise Letters}{5}{02}{L209--L216}.
\PrintBackRefs{\CurrentBib}

\bibitem [\protect \citeauthoryear {%
Lillo%
, Farmer%
\BCBL {}\ \BBA {} Mantegna%
}{%
Lillo%
\ \protect \BOthers {.}}{%
{\protect \APACyear {2003}}%
}]{%
lillo2003master}
\APACinsertmetastar {%
lillo2003master}%
\begin{APACrefauthors}%
Lillo, F.%
, Farmer, J\BPBI D.%
\BCBL {}\ \BBA {} Mantegna, R\BPBI N.%
\end{APACrefauthors}%
\unskip\
\newblock
\APACrefYearMonthDay{2003}{}{}.
\newblock
{\BBOQ}\APACrefatitle {Master curve for price-impact function} {Master curve
  for price-impact function}.{\BBCQ}
\newblock
\APACjournalVolNumPages{Nature}{421}{6919}{129--130}.
\PrintBackRefs{\CurrentBib}

\bibitem [\protect \citeauthoryear {%
Mark%
}{%
Mark%
}{%
{\protect \APACyear {2019}}%
}]{%
mark2019spoofing}
\APACinsertmetastar {%
mark2019spoofing}%
\begin{APACrefauthors}%
Mark, G.%
\end{APACrefauthors}%
\unskip\
\newblock
\APACrefYearMonthDay{2019}{}{}.
\newblock
{\BBOQ}\APACrefatitle {Spoofing and Layering} {Spoofing and layering}.{\BBCQ}
\newblock
\APACjournalVolNumPages{J. Corp. L.}{45}{}{399}.
\PrintBackRefs{\CurrentBib}

\bibitem [\protect \citeauthoryear {%
Mendonca%
\ \BBA {} De~Genaro%
}{%
Mendonca%
\ \BBA {} De~Genaro%
}{%
{\protect \APACyear {2020}}%
}]{%
mendoncca2020detection}
\APACinsertmetastar {%
mendoncca2020detection}%
\begin{APACrefauthors}%
Mendonca, L.%
\BCBT {}\ \BBA {} De~Genaro, A.%
\end{APACrefauthors}%
\unskip\
\newblock
\APACrefYearMonthDay{2020}{}{}.
\newblock
{\BBOQ}\APACrefatitle {Detection and analysis of occurrences of spoofing in the
  Brazilian capital market} {Detection and analysis of occurrences of spoofing
  in the brazilian capital market}.{\BBCQ}
\newblock
\APACjournalVolNumPages{Journal of Financial Regulation and Compliance}{}{}{}.
\PrintBackRefs{\CurrentBib}

\bibitem [\protect \citeauthoryear {%
Tuccella%
, Nadler%
\BCBL {}\ \BBA {} Serban%
}{%
Tuccella%
\ \protect \BOthers {.}}{%
{\protect \APACyear {2021}}%
}]{%
tuccella2021protecting}
\APACinsertmetastar {%
tuccella2021protecting}%
\begin{APACrefauthors}%
Tuccella, J\BHBI N.%
, Nadler, P.%
\BCBL {}\ \BBA {} Serban, O.%
\end{APACrefauthors}%
\unskip\
\newblock
\APACrefYearMonthDay{2021}{}{}.
\newblock
{\BBOQ}\APACrefatitle {Protecting Retail Investors from Order Book Spoofing
  using a GRU-based Detection Model} {Protecting retail investors from order
  book spoofing using a gru-based detection model}.{\BBCQ}
\newblock
\APACjournalVolNumPages{arXiv preprint arXiv:2110.03687}{}{}{}.
\PrintBackRefs{\CurrentBib}

\bibitem [\protect \citeauthoryear {%
Yura%
, Takayasu%
, Sornette%
\BCBL {}\ \BBA {} Takayasu%
}{%
Yura%
\ \protect \BOthers {.}}{%
{\protect \APACyear {2014}}%
}]{%
yura:2014financial}
\APACinsertmetastar {%
yura:2014financial}%
\begin{APACrefauthors}%
Yura, Y.%
, Takayasu, H.%
, Sornette, D.%
\BCBL {}\ \BBA {} Takayasu, M.%
\end{APACrefauthors}%
\unskip\
\newblock
\APACrefYearMonthDay{2014}{}{}.
\newblock
{\BBOQ}\APACrefatitle {Financial brownian particle in the layered order-book
  fluid and fluctuation-dissipation relations} {Financial brownian particle in
  the layered order-book fluid and fluctuation-dissipation relations}.{\BBCQ}
\newblock
\APACjournalVolNumPages{Physical review letters}{112}{9}{098703}.
\PrintBackRefs{\CurrentBib}

\bibitem [\protect \citeauthoryear {%
Yura%
, Takayasu%
, Sornette%
\BCBL {}\ \BBA {} Takayasu%
}{%
Yura%
\ \protect \BOthers {.}}{%
{\protect \APACyear {2015}}%
}]{%
yura:2015financial}
\APACinsertmetastar {%
yura:2015financial}%
\begin{APACrefauthors}%
Yura, Y.%
, Takayasu, H.%
, Sornette, D.%
\BCBL {}\ \BBA {} Takayasu, M.%
\end{APACrefauthors}%
\unskip\
\newblock
\APACrefYearMonthDay{2015}{}{}.
\newblock
{\BBOQ}\APACrefatitle {Financial Knudsen number: Breakdown of continuous price
  dynamics and asymmetric buy-and-sell structures confirmed by high-precision
  order-book information} {Financial knudsen number: Breakdown of continuous
  price dynamics and asymmetric buy-and-sell structures confirmed by
  high-precision order-book information}.{\BBCQ}
\newblock
\APACjournalVolNumPages{Physical Review E}{92}{4}{042811}.
\PrintBackRefs{\CurrentBib}

\end{thebibliography}

\end{document}